\documentclass[12pt,preprint]{aastex}

\begin{document}

\title{The Age and Progenitor Mass of Sirius B}

\author{James Liebert\altaffilmark{1}, Patrick A. Young\altaffilmark{2},
David Arnett\altaffilmark{1}, J. B. Holberg\altaffilmark{3}, and Kurtis
A. Williams\altaffilmark{1} }

\altaffiltext{1}{Department of Astronomy and Steward Observatory,
University of Arizona, Tucson, AZ 85721, liebert@as.arizona.edu,
darnett@as.arizona.edu, kwilliams@as.arizona.edu}

\altaffiltext{2}{Los Alamos National Laboratory, Los Alamos NM 87545
payoung@lanl.gov}

\altaffiltext{3}{Lunar and Planetary Laboratory, University of 
Arizona, Tucson AZ 85721 holberg@argus.lpl.arizona.edu}

\begin{abstract}

The Sirius~AB binary system has masses that are well determined from
many decades of astrometric measurements.  Because of the well-measured
radius and luminosity of Sirius~A, we employed the TYCHO stellar
evolution code to determine the age of the Sirius~A,B binary system
accurately, at 225--250~Myr.  Note that this fit requires the assumption
of solar abundance, and the use of the new Asplund et al. primordial
solar metallicity.  No fit to Sirius~A's position is possible using the
old Grevesse \& Sauval scale.  Because the Sirius~B white dwarf
parameters have also been determined accurately from space observations,
the cooling age could be determined from recent calculations by Fontaine
et al. or Wood to be 124$\pm$10~Myr.  The difference of the two ages
yields the nuclear lifetime and mass of the original primary star,
5.056$_{-0.276}^{+0.374}$~M$_{\odot}$.  This result yields in principle
the most accurate data point at relatively high masses for the
initial-final mass relation.  However, the analysis relies on the
assumption that the primordial abundance of the Sirius stars was solar,
based on membership in the Sirius supercluster.  A recent study suggests
that its membership in the group is by no means certain.

\end{abstract}


\section{Introduction}

The initial-final mass relation (IFMR) for progenitors and white dwarfs
is fundamental in understanding a stellar population, interpreting the
white dwarf mass and luminosity distributions, and determining the star
formation history.  Of particular interest is the upper mass limit of a
star forming a white dwarf.  It is well known that the mass loss in the
red giant phases (RGB and AGB) of stars of similar mass and chemical
composition shows dispersion.  The existence of ``horizontal branches''
in metal-poor and metal-rich clusters shows that varying amounts of the
hydrogen envelope are lost in the RGB or helium-ignition events.  For
intermediate-mass stars undergoing AGB evolution, the timing of thermal
pulses of the helium shell may produce a dispersion in the resulting
white dwarf distribution (eg. Iben \& Renzini 1983).  It may therefore
be anticipated that the the IFMR will show dispersion.

In order to determine empirically the IFMR, it is necessary to study
samples of white dwarfs where it is possible to estimate the initial
mass of the progenitor.  Since there is generally no way of establishing
the total lifetimes of white dwarfs in the field population, white
dwarfs found in well-studied star clusters have been used.  The cluster
age is known within some uncertainty, generally from fits to the main
sequence turnoff.  In a few clusters, the mass limit below which lithium
is not depleted in completely-convective low mass stars and/or brown
dwarfs may also be employed.  Several Galactic disk clusters ranging in
age from of the order 10$^8$ to 10$^9$ years, and with main sequence
turnoff masses of $\sim$2 to $>$5M$_{\odot}$, have been studied in
recent years.  These include the Hyades and Praesepe (Claver et al. 2001
is the most recent study), the Pleiades (the single white dwarf has been
studied by many authors), NGC~2516 (Koester \& Reimers 1996), NGC~3532
(Koester \& Reimers 1993), NGC~2168 (Williams, Bolte \& Koester 2004),
and NGC~2099 (Kalirai et al. 2005).  Since it is necessary to measure
the masses of the white dwarfs accurately -- by means of stellar
atmosphere fits to Balmer lines or by measuring the gravitational
redshift -- the faint white dwarf sequences recently found in the
nearest globular clusters (cf. M~4, Hansen et al. 2004) and in older 
Galactic clusters (NGC~6791, Bedin et al. 2005) are not yet as
useful.  

The upper end of the IFMR is yielding progenitors with short
nuclear-burning phases, and these lifetimes are very dependent on the
mass.  For several reasons, there must be significant uncertainty in the
estimations of the ages of such young clusters.  First, these clusters
do not necessarily have well-populated upper main sequences.  Shifts due
to rotation, unresolved binaries, and the possible presence of blue
stragglers with masses larger than the turnoff mass represent a separate
category of problems.  Finally, uncertainty in the main sequence
lifetime of stars with convective cores due to the probable overshooting
beyond the simple Schwarzschild boundary is a theoretical problem.  Thus
the uncertainty in the cluster age causes a big uncertainty in the
possible progenitor masses of white dwarfs, especially for the youngest
clusters.  These problems result in considerable uncertainty in
establishing the upper mass limit of a star that can form a white dwarf,
or even if such a uniform upper limit exists.  It is thus extremely
important if a method were available to reduce the spread in the
progenitor's mass.

Binaries consisting of a nuclear-burning star and a white dwarf offer
another potential source of candidates for the IFMR.  If the
parameters of the nuclear-burning star can be established with sufficient
accuracy that its age can be obtained by fitting its position on the HR
Diagram, then the total age of the white dwarf -- which is the sum of 
the nuclear-burning lifetime and cooling age -- can also be
established.  As for the clusters, the nuclear-burning lifetime is 
used to establish the progenitor mass of the white dwarf.  

The Sirius system is the fifth or sixth nearest stellar system to the
Sun\footnote{depending on the actual distance to the
recently-discovered, very high proper motion star SO025300.5+165258,
estimated to be within 2.4 to 3.6~pc (Teegarden et al. 2003)}, and
certainly is one of the best studied binary systems including a white
dwarf component.  One does not want to employ a white dwarf in a binary
close enough that interactions might have affected the mass loss phases
in the late stages of the progenitor's evolution.  However, the orbit is
eccentric with a period of about 50 years.  At periastron, the
components are just within 7~AU of each other (van de Kamp 1971).  This
is probably well enough separated so that any interaction during the
asymptotic giant branch phase of the original primary (now B) would have
been minimal.  Significant interaction would probably have circularized
the orbit.  We therefore assume that Sirius B evolved in a manner like
that of a single star.

An excellent trigonometric parallax, accurate to better than 1\%, is
available from the {\it HIPPARCOS} satellite mission as well as from
several good ground-based studies.  Using the bolometric energy
distribution and effective temperature determination, the luminosity of
Sirius~A is known to about 4\% accuracy.  Using the ESO {\it Very Large
Telescope Interferometer}, Kervella et al. (2003) have obtained a superb
measurement of the diameter of Sirius~A, accurate to about 0.75\%.  One
may then avoid very direct use of the T$_{eff}$ estimates of Sirius~A.
Robust parameter (T$_{eff}$, log~g) determinations for Sirius~B are also
available from space observations.  Finally, a recent reexamination of
the plethora of astrometric data on the binary orbit (Holberg 2005)
results in mass estimates of both and ``A'' and ``B'' components
accurate to 1-2\%.

Improvements in opacities, the equation-of-state, and other treatments
in modern stellar evolution codes now make possible fits to
nuclear-burning stars on the HR Diagram, especially those on or near the
main sequence (Young \& Arnett 2005).  Robust, consistent ages for
(nondegenerate) binary stars with well-determined luminosities, masses,
radii and temperature have been determined (Young et al. 2001).  

In Section 2 of this paper, we employ this code to fit the position of
Sirius~A, and estimate the age with an error estimate.  This gives the
systemic age from which to subtract the cooling age of ``B'' in Section
3 to obtain the estimate of its progenitor mass, and its uncertainty.
We then summarize and discuss an important caveat in Section 4. 

\section {The Fitting of Sirius~A in an HR-like Diagram}

The mass adopted for Sirius~A is 2.02$\pm$0.03M$_{\odot}$ (Holberg 2005).
Note that this new astrometric study yields a mass about 5\% smaller
than some earlier determinations of Sirius~A.  (The smaller orbit also 
yields a smaller mass for Sirius~B.)  The adopted radius of Sirius~A 
is 1.711$\pm$0.013~R$_{\odot}$ (Kervella et al. 2003), and the 
resulting luminosity is L/L$_{\odot}$ = 25.4$\pm$1.3.  Since the 
luminosity and radius are more directly measured than the T$_{eff}$, 
these will be the quantities fitted in an $L$--$R$ equivalent of the 
HR diagram.  

One issue with Sirius~A is that it is a chemically-peculiar A1V dwarf,
so that a direct determination of the interior chemical composition is
not possible.  However, it has been believed for a long time that Sirius
is a member of a large moving group near the Sun, called the ``Sirius
supercluster'' (Hertzsprung 1909; Eggen 1983).  (We reassess this issue 
in the last section of the paper, however.)  The ``core'' of this
association is the Ursa Major ``dipper'' stars.  Metallicity estimates
for the group members (that are not chemically-peculiar) are generally
consistent with solar (Palous \& Hauck 1986).  Eggen (1992) compared the
abundances of the Sirius and Hyades superclusters and concluded that the
former is deficient by about -0.18 dex in [Fe/H] compared with the
Hyades group.  The latter is believed to exceed solar by about this
amount.  There is no guarantee that appreciable dispersion in
metallicity does not exist among the group.  We nonetheless can do no
better than assume solar X,Z values below.

However, a new solar abundance scale determined from 3D non-LTE
calculations of the solar atmosphere (Asplund et al. 2004, and
references therein) has significantly lower abundances of oxygen, carbon
and nitrogen.  The overall heavy elements abundance parameter for the
Sun decreases to a primordial value of Z=0.014.  The new scale
jeopardizes the excellent agreement of the ``standard solar model'' with
the helioseismology observations (Bahcall et al. 2005), though a new
study attempts to remedy this, retaining the Asplund et al. values
(Young \& Arnett 2005).  We have adopted the new scale for the
calculations described below.

In Figure~1 are shown the results of running the TYCHO code (Young \&
Arnett 2005) to fit the position of Sirius~A.  The model included
wave-driven mixing and diffusion. The former is a way of accounting for
the ``overshooting'' of the convective core based on physics, rather
than a prescription involving the ratio of the mixing length to scale
height.  The microphysics for opacities, reaction rates and the EOS
appear numerically similar to those of Ventura et al. (1998).  The
inclusion of these effects produces only a small effect at the best fit
age of 237.5$\pm$12.5 Myr. Radiative levitation was not included.  A perfect
fit would require a lower metallicity or an increase in mass of a few
hundredths M$_{\odot}$. The evolution was begun well up the Hayashi
track and followed throughout the pre-main sequence. The best-fit age
includes the pre-MS evolution. No pre-MS accretion was included. The
best-fit age is constrained primarily by the radius determination.

The precision with which the Sirius-A radius and luminosity are known is
such that the use of the new solar abundance scale matters greatly.  In
particular, {\it no reasonable fit can be achieved} if the old scale
(Grevesse \& Sauval 1998) were adopted.  When the 2.02M$_{\odot}$ track
for the old solar scale crosses the correct radius on its main sequence
track, the luminosity is log L/L$_{\odot}$ = 1.26, several sigma below
the observed value.  The new solar abundances result in lower interior 
opacities, so the star has a smaller radius at a given luminosity, and 
vice versa. 

Another estimate of the radius of Sirius~A is reported by Decin et al.
(2003) using the Short Wavelength Spectrometer on the Infrared Space
Observatory.  This measurement is far less precise, and the Kervella et
al. (2003) angular diameter is within its 5\% errors.

\section{The cooling age and progenitor mass of Sirius~B}

An accurate T$_{eff}$ determination is also essential for measuring the
cooling age of Sirius~B.  Holberg et al. (1998) employed space
ultraviolet {\it Extreme Ultraviolet Explorer} and {\it International
Ultraviolet Explorer} spectrophotometry to estimate 24,790$\pm$100.
Barstow et al's (2005) estimate, using an optical spectrum of
extraordinarily high signal-to-noise ratio, obtained with the Space
Telescope Imaging Spectrograph ({\it STIS}) on the {\it Hubble Space
Telescope} is 25,193$\pm$37 (an internal error only).  We adopt
T$_{eff}$ = 25,000$\pm$200.

The astrometric reanalysis of Holberg (2005) results in a measurement of
1.00$\pm$0.01M$_{\odot}$ for the mass of Sirius~B.  Barstow et
al. (2005) obtained T$_{eff}$ = 25,193~K (as stated above) and log~g =
8.528$\pm$0.05 from fits to the Balmer line profiles.  In the next
paragraphs we shall apply cooling sequences to obtain the cooling age of
Sirius B.  These also yield a second relationship (besides the surface
gravity) between the radius and mass -- R = R(M,T$_{eff}$).  Solving for
the mass one obtains 0.978$\pm$0.005M$_{\odot}$, if the cooling
sequences of Wood (1995) are used, or 0.003M$_{\odot}$ less if the those
of Fontaine et al. (2001) are employed.  Using their gravitational
redshift measurement of 80.42$\pm$4.83 km~s$^{-1}$ -- yielding the ratio
M/R -- with the Wood sequences, Barstow et al. (2005) obtain
1.02$\pm$0.02M$_{\odot}$. Note that these mass determinations are
generally below previous estimates in earlier literature.  For this
study, we shall adopt 1.00$\pm$0.02M$_{\odot}$ for Sirius B.

For the cooling ages of white dwarfs, most cluster white dwarf studies
in the last ten years have used the evolutionary models of Wood (1992,
1995).  For reasons stated below, we will use primarily a new sequence
by Fontaine, Brassard, \& Bergeron (2001).  Both calculations used 50\%
carbon -- 50\% oxygen cores, and the relatively thick outer layers of
helium and hydrogen predicted by most evolutionary calculations of the
asymptotic giant branch phase.

The Fontaine et al. calculations incorporate several new treatments of
physics (Fontaine 2005, private communication).  First, the equation of
state of Saumon, Chabrier, \& Van Horn (1995) for H and He in the
partially-ionized envelope, and a new treatment of carbon in the same
regime, was employed.  Second, the conductive opacity tables of Hubbard
\& Lampe (1969) and Itoh \& Kohyama (1993, and references therein) were
``fused together,'' rather than treated as completely separate.  Third,
no discontinuities in the chemical composition profile are allowed.
Diffusion at the interfaces is calculated.  Fourth, a full stellar
structure is evolved from the center to the top of the atmosphere, not a
simple ``Henyey'' core which excludes the envelope.  Fifth, a robust and
accurate finite element technique (developed by P. Brassard) is used, as
opposed to finite differences as in most other codes.  We thus have
adopted these sequences to estimate the cooling age of Sirius~B.  Having
said this, it is interesting and reassuring that the 1.00~M$_{\odot}$
C,O sequence of Wood (2005, private communication) reaches a cooling age
of 123.3 Myr at 25,000~K, while the Fontaine et al. (2001) sequence with
identical parameters reaches 123.6$\pm$10~Myr.

Error bars in the ``final'' mass of Sirius~B are as stated above.  The
uncertainty in the mass and systemic age (\S~2) leads to a nuclear 
lifetime for the progenitor in the range of 101--126 Myr.  

For determining the initial mass, we consider errors due to (1)
uncertainty in the final mass, and (2) uncertainty in the systemic age
(\S~ 2).  We shall also consider a possible additional uncertainty due
to the different results obtained from the TYCHO and ``Padova'' codes,
and an uncertainty due to the unknown carbon-oxygen abundance profiles
throughout the core.

For most recent papers on white dwarf sequences, such as Ferrario et
al. (2005, hereafter F05), the ``Padova'' stellar evolution models of
Girardi et al. (2002) were used to get the nuclear lifetime and the
initial progenitor mass.  For this paper, for self-consistency with
\S~2, we present first the calculations of these using the TYCHO code.
 
While TYCHO incorporates physical treatments available in 2005 (rather
than several years earlier for the ``Padova'' literature calculations),
the principal difference between the two results may be the treatment of
mixing beyond the traditional Schwarzschild core boundary.  Girardi et
al. (2002) use an ``overshoot'' prescription based on a single
parameter.  As alluded to in \S~2, TYCHO includes the effects of
hydrodynamics in the convective boundary and radiative regions in a
predictive, physically-motivated fashion.  This treatment may give more
accurate predictions of core sizes, and thus of luminosities, radii, and
nuclear lifetimes.  

From the TYCHO calculations, the masses that bracket the above range in
the progenitor's nuclear lifetime are 5.43 and 4.78~M$_{\odot}$, with a
mean of 5.056M$_{\odot}$.  The uncertainly in the cooling age (due to
that of the white dwarf mass) contributes errors of only -0.171 and
+0.262M$_{\odot}$.  The uncertainty in the age of the binary system
contributes -0.213 and +0.273M$_{\odot}$.  When added in quadrature,
these values yield the mass range of the previous paragraph.  If we
employ instead the Girardi et al. (2002) tables, again for nuclear
lifetimes of 101 Myr and 126 Myr, the mean value of initial mass is only
slightly larger at 5.132$_{-0.23}^{+0.28}$M$_{\odot}$.  Thus we may also
conclude that the physical treatment of the convective boundary region
by TYCHO, and the ``overshoot'' prescription of the Padova group, agree
pretty well near 5M$_{\odot}$.  In any case the uncertainties in the
initial mass from this analysis are smaller than for any massive cluster
white dwarf included in F05.

An additional source of uncertainty in the cooling age not considered
here is that due to the carbon--oxygen abundance distribution in the
core.  As stated previously, the white dwarf cooling calculations of
Fontaine et al. (2001) simply employ a 50\%--50\% mixture.  The usual
practice in the published analyses of the cluster white dwarfs is, as
stated previously, to use Wood's calculations with the same mixture.
Other available calculations from Wood are for pure carbon and for pure
oxygen compositions.

The 5~M$_{\odot}$ sequence calculated from TYCHO produces very nearly a
1M$_{\odot}$ core at the end of the AGB evolution, which suggests that
the prescription for mass loss in the red giant phases is fairly
accurate.  The resulting white dwarf has a carbon abundance of 15\%
throughout the inner 0.45~$M_{\odot}$, comprised of the former
convective He-burning core. The carbon abundance increases from 35\% to
40\% moving outward through the region processed by He shell burning,
with a narrow ($\sim 0.01 M_{\odot}$) spike of 60\% carbon where He
burning is incomplete. At low He abundances The $Y_{\rm ^4He}^3$
dependence of the triple $\alpha$ reaction favors the ${\rm
^{12}C(\alpha,\gamma)^{16}O}$ reaction over triple $\alpha$. More
massive cores with higher entropy favor the production of $^{16}O$ over
$^{12}C$. Convective He burning cores also tend to grow at low $Y_{\rm
^4He}$ due to increased opacity.  This process as well as any
non-convective mixing process which mixes He into the core at low
abundance will increase the destruction of $^{12}C$, resulting in an
oxygen dominated core (Arnett 1996). Pure carbon white dwarf cooling
models are thus excluded for this mass range, favoring a shorter cooling
time.  Using the simple mixture is therefore the best choice of
available cooling sequences, but the additional error to the cooling
time due to not matching the true abundance profile is not well
constrained.

\section{Conclusions and a Caveat} 

The result of this analysis is a spread in the possible progenitor mass
of Sirius~B generally smaller than those in the young clusters cited
above.  In a parallel F05 paper, Sirius~B provides a valuable ``anchor
point'' in the high mass part of the IFMR, which is plotted therein
(their Fig.~1).  

The conclusions as to the Sirius~B progenitor mass obviously depend on
the fit to the Sirius~A luminosity and radius being a valid measure of
the systemic age.  The ``Achilles heel'' of the analysis may be the
assumption of solar composition for the primordial abundance of the two
stars.  Since there is no way to measure directly the primordial
abundance of either the chemically-peculiar A star or the white dwarf,
we have relied on the assumption that it is a member of the Sirius
supercluster, and that these stars have generally been shown to have
abundances indistinguishable from solar.  It must be acknowledged that,
in the recent study of King et al. (2003), it appears by no means
certain that Sirius -- so distant from the Ursa Major core -- is a
member of the Sirius supercluster.  These authors estimate an age of
500$\pm$100 Myr for the supercluster from main sequence turnoff fits.
Using Str\"omgren photometry, Asiain et al. (1999) estimate 520$\pm$160
Myr.  These values are substantially larger than previous literature
estimates -- eg. 240 Myr (Eggen 1983).  They are appreciably larger than
the systemic age of 225--250~Myr determined in this paper for Sirius.
There is no way, if this analysis is based on sound assumptions, that
the Sirius age can be 400~Myr.  King et al. (2003) do emphasize that
they have not determined that the supercluster is coeval, nor that the
stellar abundances are the same.

If we were to consider the possibility that the interior abundances of
Sirius A are not determined, we consult the comprehensive study by
N\"ordstrom et al. (2004) of 7566 nearby, single F and G dwarfs to see
what the abundance distribution is.  For the subsample within 40 pc of
the Sun, and at estimated ages near 1 Gyr or less, the [Fe/H]
determinations (see their Fig.~28) range from about -0.3 to +0.2, with a
mean value near -0.02 to -0.06 (the former if the sample is restricted
to the subsample with age estimates considered to be accurate).  As
mentioned in \S~2, the fit would actually improve (though its already
within one sigma), if the assumed metallicity (Z) were decreased.  At
the extreme, with Z = 1/2 solar, a good fit can be achieved at an age
of 375$\pm$19~Myr, over 50\% larger.  (The metal-poor star begins with a
smaller radius and higher T$_{eff}$, and must evolve farther through its
main sequence phase to reach the observed radius and luminosity.)  This
systemic age would correspond to a much lower progenitor mass of
3.61$\pm$0.125$M_{\odot}$.  This unlikely outcome would make Sirius 
a seriously-discrepant data point in the IFMR (F05). 

On the other hand, we may note that most stars in the solar
neighborhood have close to solar abundances.  The analysis in this
paper is self-consistent within the stated assumptions.  As a point in
the overall IFMR for disk stars, it can be seen in the F05 paper that
the solar abundance Sirius~B point has a somewhat higher white dwarf
mass or a somewhat lower initial mass than most of the similar white
dwarfs in NGC~2516, M35 and the Pleiades, but overlaps the error bars
of most of these.  Marigo (2001) gives a careful treatment of AGB mass
loss, and predicts that a solar metallicity 5.06$M_{\odot}$ star
should produce almost exactly a 1$M_{\odot}$ white dwarf.  (We
remarked earlier that the TYCHO code does also.)  In summary, this
result appears to provide a strong confirmation of stellar theory.

\acknowledgments

This work was supported by the National Science Foundation through grant
AST-0307321 (JL and KAW).  We thank Gilles Fontaine, Pierre Bergeron,
Martin Barstow, and Matt Wood for valuable communications, Eric
Mamajek for a tutorial on moving groups, and the anonymous referee for 
several helpful suggestions.

\newpage

\begin{figure} 

\caption{The log luminosity vs. log radius TYCHO evolutionary track
beginning with an underluminous starting pre-main sequence model.  The
position of Sirius~A with error bars is shown and is fit by the track
within one sigma at a total (pre-ms + ms) age of 237.5$\pm$12.5 Myr,
about 1/7 of its main sequence lifetime. }  

\end{figure}

\end{document}